\documentclass [prb,twocolumn,notitlepage,superscriptaddress,floatfix] {revtex4-1}
\usepackage{amsmath}
\usepackage{graphicx}
\usepackage{lmodern}
\usepackage{amsmath}

\usepackage{color}
\usepackage{amssymb}
\newcommand{\beq} {\begin{equation}}
\newcommand{\eeq} {\end{equation}}
\newcommand{\bea} {\begin{eqnarray}}
\newcommand{\eea} {\end{eqnarray}}
\newcommand{\be} {\begin{equation}}
\newcommand{\ee} {\end{equation}}
\renewcommand{\(}{\left(}
\renewcommand{\)}{\right)}
\renewcommand{\[}{\left[}
\renewcommand{\]}{\right]}

\DeclareMathOperator{\sgn}{sgn}

\begin{document}

\title {Enhancement of superconductivity at the onset of charge-density-wave order in a metal}
\author{Yuxuan Wang}
\affiliation{Department of Physics, University of Wisconsin, Madison, WI 53706, USA}
\author{Andrey V. Chubukov}
\affiliation{William I. Fine Theoretical Physics Institute,
and School of Physics and Astronomy,
University of Minnesota, Minneapolis, MN 55455, USA}
\date{\today}

\begin{abstract}
We analyze superconductivity in the cuprates near the onset of an incommensurate charge density wave (CDW) order with momentum ${\bf Q} = (Q,0)/(0,Q)$, as observed in the experiments.  We first consider a semi-phenomenological charge-fermion model in which hot fermions, separated by ${\bf Q}$, attract each other by exchanging soft CDW fluctuations.
 We find that
  in a quantum-critical region near CDW transition, $T_c = A {\bar g}_c$, where ${\bar g}_c$ is charge-fermion coupling and $A$ 
  is the prefactor which we explicitly compute. We then consider the particular microscopic scenario in which CDW order parameter emerges as a composite field made out of primary spin-density-wave fields.
 We
 show that charge-fermion coupling ${\bar g}_c$ is of order of spin-fermion coupling ${\bar g}_s$. As the consequence,
   superconducting $T_c$
   is substantially enhanced near the onset of CDW order.
Finally we analyze the effect of an external magnetic field $H$.
 We show that, as $H$ increases,
   optimal $T_c$ decreases and the superconducting dome becomes progressively more confined to the CDW quantum-critical point.
These results are
 consistent with the experiments.
 \end{abstract}
\maketitle

\section{Introduction}
Understanding of the nature of charge order in high-$T_c$  cuprates and of its effect on superconductivity is essential for the full understanding of the complex physics
in this materials.
 Charge order has been observed in the cuprates quite some  time ago~\cite{tranquada,tranquada1}, but was originally  though to be present only in  La-based materials.  Recent wave of discoveries of an incommensurate charge-density-wave (CDW) order in Y-, Bi-, and Hg- based cuprates~\cite{ybco,ybco_2,ybco_1,X-ray, X-ray_1,davis_1,mark_last,greven} has demonstrated that
 charge order is ubiquitous to all families of high-$T_c$ cuprates.
 A true long-range CDW order has so far been observed only
 in a finite magnetic field, but short-range static order (probably pinned by impurities) exists already in zero field.  On a phase diagram, CDW order
  been detected within the pseudogap region, and its onset temperature is the highest around  doping level $x \sim 0.12$.
The CDW has
 an incommensurate
  momentum ${\bf Q}=Q_y=(0,Q)$ or $Q_x=(Q,0)$ and the order is likely uni-axial, i.e., it develops, within a  given domain, with only $Q_x$ or $Q_y$ (Ref.\ \onlinecite{davis_1}).

 The discovery of the CDW order raised a number of questions about its origin~\cite{ms,efetov, laplaca,charge,patrick,pepin,debanjan,flstar,kivelson,atkinson,charge_new}, a potential  discrete symmetry breaking
  before a true CDW order sets in~\cite{ybco_2,kerr,sidis,nie,charge,tsvelik,agterberg,atkinson_new,taillefer_nem}, and the
  relation between  CDW order (or its fluctuations)  to pseudogap behavior~\cite{charge,atkinson,atkinson_new,flstar} and Fermi surface (FS) reconstruction~\cite{FS_reconstruct}.

  In this paper we discuss another issue related to CDW -- its
   effect on superconductivity.  We take as inputs three experimental
    facts.
   First, $T_c$, as a function of doping, has a dip or a plateau at around  $x \sim 0.12$, where the
    onset temperature of CDW is the
     largest \cite{18lbco,18ybco}.
    Second, when CDW is suppressed by applying pressure~\cite{taillefer_pressure}, superconducting $T_c$
       increases. Third, when a magnetic field is applied~\cite{ramshaw}, the dip grows and at high enough field  the superconducting dome splits into two,
       and the one at larger dopings is centered at the same $x$  at which
        CDW order
         develops at $T=0$ (Ref.\ \onlinecite{ramshaw}).  In other words, superconductivity forms a dome on top of quantum-critical point (QCP) for the onset of the CDW order.

  The first two results can be naturally understood if we assume that CDW and $d$-wave superconductivity are just competing orders, i.e., when one order is at its peak, the other one is suppressed.  The third observation, on the other hand, requires one to go beyond a simple ``competing order" scenario because the presence of the dome of superconductivity on top of CDW  QCP  indicates that superconductivity is at least partly caused by soft fluctuations of CDW order which then must
   develop at larger energies than
    the ones related to superconductivity.

   In our analysis we explore CDW-mediated superconductivity in some detail.  We perform our analysis in two stages.
    At the first stage we put aside the issue what causes CDW order, assume that this order develops below some critical doping,
       and consider a semi-phenomenological model of fermions interacting by exchanging soft CDW fluctuations with momenta ${\bf Q}$.
     This model is quite similar to
      the spin-fermion model, considered in earlier studies of spin-mediated superconductivity
       for cuprates, Fe-pnictides, and other correlated materials~\cite{cuprates,benfatto,cps,chi,rmp,am_subir},
        and we dub this model the ``charge-fermion model".
       The  charge-fermion and spin-fermion models are similar but  differ in detail because of the difference between the CDW momentum ${\bf Q}$ and antiferromagnetic momentum $(\pi,\pi)$,
       and also because of the difference in the spin structures of charge-mediated and spin-mediated interactions (spin Kronecker $\delta$ functions vs spin Pauli matrices).
      One qualitative consequence of these differences is that charge-mediated interaction gives rise to superconductivity with
       a full gap in each region
       where the FS crosses the Brillouin zone boundary (anti-nodal regions in the cuprate terminology), but it does not couple superconducting order parameters from different regions, hence it alone cannot distinguish between $s$-wave and $d$-wave
       pairing symmetries~\cite{italians}.

        We perform quantitative analysis of the pairing within the charge-fermion model in the most interesting quantum-critical  regime right  above QCP for CDW order.  In this regime, charge-mediated interaction gives rise to the pairing but also destroys coherence of hot fermions (the ones at the FS separated by ${\bf Q}$). Superconducting $T^{\rm ch}_c$ is determined by the interplay between strong tendency towards pairing and strong pair-breaking effects associated with the self-energy.  We compute Landau damping of soft bosons and fermionic self-energy, and then
         obtain and solve the linearized gap equation with renormalized fermionic and bosonic propagators.  We show that in the quantum-critical region $T_c$ is finite and scales with the effective charge-fermion coupling constant ${\bar g}_c$: $T^{\rm ch}_c = A_c {\bar g}_c$, where
          $A_c \approx 0.0025$.

    At the second stage we consider the specific microscopic scenario for CDW order -- the one in which CDW is a composite
     order parameter made out of primary $(\pi,\pi)$ spin fluctuations. Within this scenario, spin fluctuations are assumed
     to develop first, at energies comparable to bandwidth, while CDW fluctuations develop at smaller energies and do not provide substantial feedback on spin fluctuations. This composite order scenario requires fermion-fermion interaction  to be comparable to the bandwidth (otherwise spin fluctuations do not develop at high energies) and inevitably is partly phenomenological.  We will not discuss a complementary, renormalization group-based, truly weak coupling scenario in which all fluctuations (spin, charge, superconducting) develop simultaneously at low energy and mutually affect each other~\cite{rg}.

      Various versions of magnetically induced charge bond and site orders have been proposed over the last few years~\cite{ms,efetov, laplaca,charge,pepin,debanjan,flstar,atkinson,charge_new,greco,norman_last}, some focused on CDW
       with diagonal momentum $(Q,Q)$, and others on CDW
        with  momenta  $Q_x =(Q,0)$ and $Q_y =(0,Q)$.
       Motivated by experiments, we 
base our discussion on the soft fluctuations of
 axial CDW
       with momenta $Q_x$ or $Q_y$.
        Previous studies have found~\cite{coex}
        that axial CDW has a partner -- an incommensurate pair-density wave (PDW), and fluctuations in CDW and PDW channels
        develop simultaneously.
       To keep presentation focused, we concentrate on CDW and neglect PDW fluctuations.  The latter
         can in principle also mediate pairing interaction, but are unlikely to destructively interfere with CDW fluctuations.

\begin{figure}
\includegraphics[width=\columnwidth]{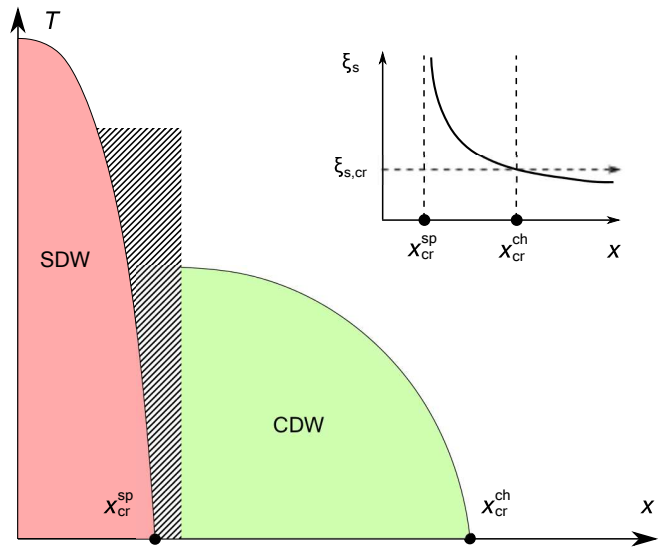}
\caption{Doping range where CDW order with momentum $(Q,0)$ or $(0,Q)$  emerges within itinerant spin-fluctuation scenario.
 The effective interaction in the CDW channel is made out of two spin-fluctuation propagators.  The CDW  order develops only
when the system is sufficiently close to a magnetic instability and terminates at doping $x^{\rm ch}_{\rm cr}$, different from $x^{\rm sp}_{\rm cr}$ for
 antiferromagnetism.  In the shaded region near $x^{\rm sp}_{\rm cr}$  localization of electronic states (Mott physics) becomes relevant 
 (Ref.\ \onlinecite{mott}),
  and
   spin-fluctuation approach needs to be modified. In this region, the onset temperature of CDW order drops.}
\label{phases_cdw}
\end{figure}

    In an earlier work~\cite{charge} we have shown that
    the axial CDW order develops in a paramagnet at a
    finite $T_{\rm cdw}$,  provided that  magnetic correlation length $\xi_s$ is large enough. As $\xi_s$  decreases,  $T_{\rm cdw}$ also decreases and vanishes at some finite $\xi_{\rm s,cr}$, setting up a CDW QCP at a finite distance from a magnetic QCP (the one at $\xi_s = \infty$).  Near $\xi_{s, {\rm cr}}$ CDW fluctuations become soft and give rise to singular pairing interaction mediated by these fluctuations.
    We show
     the behavior of $T_{\rm cdw}$ vs  $\xi_s$ schematically in Fig.\ \ref{phases_cdw}.

    Whether this additional pairing interaction substantially affects superconducting $T_c$ depends on the interplay
      between charge-fermion coupling ${\bar g}_c$ and the coupling ${\bar g}_s$  between fermions and primary spin fluctuations.
      The argument is that
      spin-fluctuation exchange by itself gives rise to
       superconducting pairing, and
       at large $\xi_s$ the corresponding $T^{\rm sp}_c$ scales with ${\bar g}_s$:
     $T^{\rm sp}_c = A_s {\bar g}_s$, where
     $A_s \approx 0.007$ (Ref.\ \onlinecite{wang}).
     As $\xi_s$ decreases, $T^{\rm sp}_c$ also decreases but remains finite.
      The two-dome
       structure
       of $T_c (x)$,
        observed in the cuprates in an applied magnetic field, can be understood within this approach only if near $\xi_s = \xi_{\rm s,cr}$, $T^{\rm ch}_c \geq T^{\rm sp}_c$. If this does not hold, i.e.,  $T^{\rm ch}_c$ is smaller than $T^{\rm sp}_c$, the contribution to superconductivity from charge-mediated exchange is only subdominant to the one from spin fluctuations. In this situation, the only effect on $T_c$ from CDW is due to direct competition between CDW and superconducting orders.
          This competition may give rise to  additional reduction of $T_c$ in a magnetic field, given the experimental evidence that
            CDW order increases in a field.  However, it cannot give rise to a peak of $T_c$ above CDW QCP.

We evaluate charge-fermion coupling ${\bar g}_c$  within the RPA-type analysis of charge fluctuations near CDW QCP:
$U_c^{\rm eff} (q) \propto U_c/(1-U_c \Pi_c (q)) \equiv {\bar g}_c/(\xi^{-2}_c + (q-Q)^2)$,
 and show that
$\bar g_c$
  is comparable to spin-fermion coupling ${\bar g}_s$. This result may look strange
   because charge fluctuations, viewed as composite
     objects made out of spin fluctuations, develop only in a narrow range near the FS points separated by $Q_x$ or $Q_y$, with the width in momentum space of order $\Lambda\sim\xi_s^{-1}$.  As the consequence, ${\bar g}_c \sim U_c \xi_s^{-2}$. However, the
  ``bare" interaction in the charge channel, $U_c$, is a composite object made out of two spin fluctuation propagators and two fermionic Green's functions (see Fig.\ \ref{fig:x}). This composite object behaves as ${\bar g}_s \xi_s^2$. As the consequence, $U_c \xi_s^{-2}$ is not reduced by $\xi_s$, and
    ${\bar g}_c$ differs from ${\bar g}_s$ only by a numerical factor.

 To properly calculate the ratio ${\bar g}_c/{\bar g}_s$ one needs to do full-scale dynamical calculations, even if one restricts with ladder series of diagrams, like in RPA.  In this work, we use a simplification and approximate the bare interaction in the charge channel $U_c$ by a constant within  the
  momentum  range $\Lambda \sim \xi_s^{-1}$ around proper Fermi surface points (hot spots) and set it to zero outside this range.  We compute the polarization operator $\Pi_c (q, \Omega_m)$ and use RPA to obtain charge-mediated effective interaction within fermions.  We use the condition for CDW QCP: $1 = U_c \Pi_c (Q)$ to fix $\Lambda$ and obtain explicit relation between ${\bar g}_c$ and ${\bar g}_s$. Within this approach, we find
   ${\bar g}_c \geq {\bar g}_s$.

\begin{figure}
\includegraphics[width=\columnwidth]{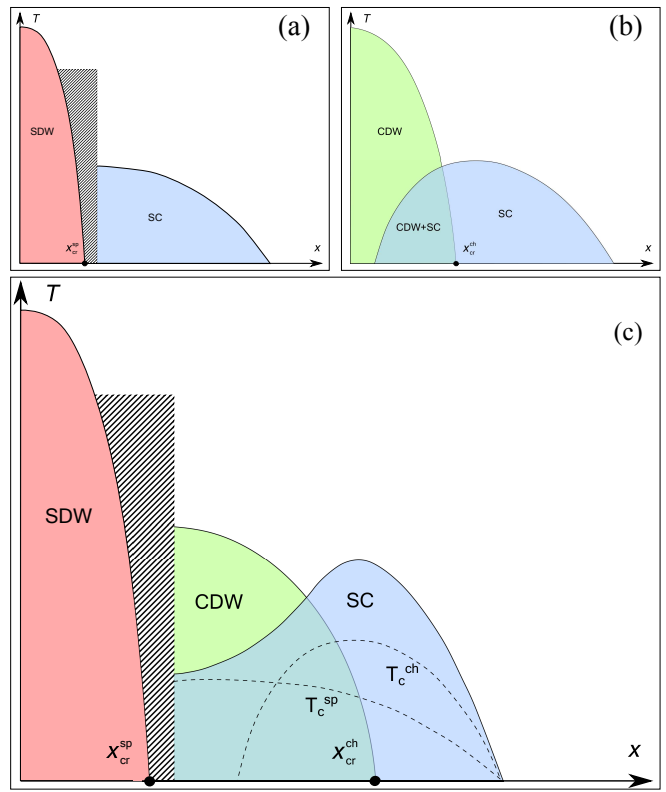}
\caption{Schematic phase diagram for the interplay between superconductivity mediated by spin and charge fluctuations.
Panel (a): the onset temperature $T_c^{\rm sp}$ of spin-mediated superconductivity, as a function of doping. We assume that impurities kill superconductivity above a certain doping.
Panel (b): the onset temperature $T_c^{\rm sp}$ of charge-mediated superconductivity near the onset of CDW order. We compute  $T_c^{\rm ch}$ in this work. Panel (c): The full phase diagram.  The non-monotonic $T_c (x)$ is obtained by combining spin-mediated and charge-mediated contributions to $T_c$ from panels (a) and (b) (dashed lines). }
\label{phases}
\end{figure}

We argue, based on this analysis, that the enhancement of superconductivity near a CDW QCP is substantial, i.e superconductivity in the cuprates
 comes from both spin and charge fluctuations.
   We present the schematic phase diagram in Fig.\ \ref{phases}.
   This scenario  explains the developments of two domes of $T_c$ in a high magnetic field: one, at smaller doping, is due to critical spin fluctuations, and another, at larger doping, is
    due to critical charge fluctuations.

\subsubsection{Relation to earlier works}

The pairing by charge fluctuations has  been studied before.  In the context of the cuprates, DiCastro, Castellani, Grilli, and their collaborators~\cite{italians} analyzed in detail the pairing mediated by axial CDW fluctuations near the onset of  charge order, which was assumed to develop on its own rather than being induced by SDW fluctuations  They found that charge-mediated 4-fermion interaction is attractive in both $d$-wave and $s$-wave channel and does not distinguish the two. They argued that some other mechanism, e.g., antiferromagnetic spin fluctuations, lifts the degeneracy in favor of $d$-wave.  We obtain the same results in Sec. II. The novel part of our analysis in this Section is the calculation of charge-mediated $T_c$ in the quantum-critical regime.

This pairing problem near CDW QCP  has certain similarities to the pairing  at the onset of a nematic order, which has been extensively studied in recent years~\cite{max_new, senthil, sam, sslee}.
 It has been argued that ${\bf Q} =0$ nematic fluctuations enhance all partial components of the pairing susceptibility.  The case of QCP at small $Q_x/Q_y$ is less unrestrictive in this respect but still, $s$-wave and $d$-wave channels are degenerate for CDW fluctuations.

 The analysis of the electron-mediated pairing near a QCP for density-wave order is also quite interesting from a general perspective as it adds one important new element not present for the pairing away from a QCP. Namely, the same interaction which gives rise to strong attraction also destroys
  fermionic coherence and prevent fermions from developing supercurrent~\cite{nfl,acs}.  Superconductivity then may or may not  emerge depending on the  interplay between these two opposite tendencies~\cite{max_new,extra}.

The rest of this paper is organized as follows. In Sec.\ II we
 introduce and analyze  charge-fermion model of itinerant electrons coupled to near-critical CDW fluctuations. In Sec.\ IIa  we derive
 bosonic
 and
 fermionic self-energies.
 In Sec.\ IIb we study the pairing problem and
  obtain $T^{\rm ch}_{c}$ due to charge-fluctuation exchange near a CDW QCP. We show that $T^{\rm ch}_{c}$ scales with the charge-fermion coupling constant
${\bar g}_c$.
    In Sec.\ III, we
     relate
${\bar g}_c$ and spin-fermion coupling ${\bar g}_s$
  within the magnetic scenario for CDW. In this scenario,
    a CDW order parameter field emerges as a composite object made out of two spin-fluctuation propagators.
  We show that ${\bar g}_c$ is comparable to
   ${\bar g}_s$ and
   may
    even exceed
     it.
In Sec.\ IV we discuss in some more detail superconducting dome around CDW QCP.  Finally in Sec.\ V we discuss the results and present our conclusions.

\section{The charge-fermion model}
\begin{figure}
\includegraphics[width=\columnwidth]{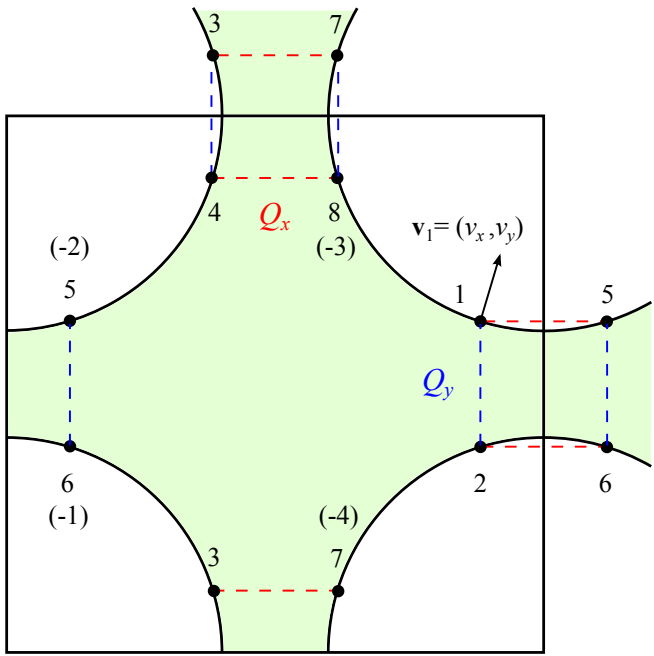}
\caption{The Fermi surface of a 2D electronic system on a square lattice and location of hot spots for charge-mediated interaction.
 CDW ``hot" spots are defined as points on the Fermi surface  separated by either $Q_y=(0,Q)$ or by $Q_x=(Q,0)$.
 In a generic situation, there are 8 hot spots for $Q_y$ and eight for $Q_x$. Motivated by experiments, we consider the case
 when hot spots for $Q_y$ and  for $Q_x$ merge. In this situation, CDW hot spots coincide with hot spots for $(\pi,\pi)$ magnetism, and there are
 eight of them on the Fermi surface We label  these hot spots by 1-8 ($5 \equiv -2$, $6 \equiv -1$, etc). The arrow shows the direction of the Fermi velocity at hot spot 1. Fermi velocities at other seven hot spots are related to this one by symmetry.
Near CDW instability hot fermions interact with each other by exchanging CDW fluctuations with momenta $Q_{x,y}$. Here and in other figures below
 we represent charge fluctuations by dashed lines. }
\label{fig:4}
\end{figure}
We begin with a semi-phenomenological analysis.  We assume, without specifying the reason,  that CDW order with momentum $Q_y=(0,Q)$ and/or $Q_x=(Q,0)$,
  develops at some critical doping $x_c^{\rm cr}$, and consider the model of 2D fermions interacting by exchanging near-critical, soft charge fluctuations.
   We dub this model as ``charge-fermion model" by analogy with the spin-fermion model which was introduced to describe a system near a magnetic QCP.
    As our goal is to describe low-energy physics (energies well below the bandwidth), we focus on momentum ranges around the CDW ``hot" spots on the FS, defined as points which are separated by CDW momentum $Q_x$ or $Q_y$.   For a generic ${\bf Q}$  there  are 16 CDW hot spots, eight corresponding to $Q_y$  and another eight corresponding to $Q_x$. For simplicity we consider the case when ${\bf Q}$ is such that CDW hot spots are at or near the crossing between the FS and symmetry lines in the Brillouin zone $k_x \pm k_y = \pm \pi$. Then hot spots from $Q_y$ sector merge with hot spots from $Q_x$ sector, and the total number of hot spots become eight.
    We label these points as 1-8 in Fig.\ \ref{fig:4}.  This approximation works reasonably well for the values of ${\bf Q}$ extracted from the experiments~\cite{X-ray}.

 We  linearize the fermionic dispersion in the vicinity of a hot spot $i$  as $\epsilon_{i,\tilde k}={\bf v}_i\cdot \tilde{\bf k}_i$, where ${\bf v}_i$ is the Fermi velocity  and $\tilde {\bf k}_i$ is the momentum deviation from the hot spot $i$. We define the Fermi velocity at the hot spot 1 as ${\bf v}_{1}=(v_x,v_y)$, the velocities at other hot spots
   are related by symmetry.
    In the cuprates, the FS at the antinodal region is ``flattened", and we have $v_x<v_y$ (see Fig.\ \ref{fig:4}). We define $\alpha=v_x/v_y< 1$.

The action of the charge-fermion model can be written as
\begin{align}
\mathcal{S}=&\int d\tilde k\sum_{i,\alpha} c_{i\alpha}^\dagger(\tilde k)(-i\omega_m+\epsilon_{i\tilde k})c_{i,\alpha}(\tilde k) \nonumber\\
&+\int d\tilde q~\chi_{0c}^{-1}(\tilde q)\sum_{a=x,y}\phi_a(\tilde q)\phi_a^{\dagger}(\tilde q) \nonumber\\
&+g_c\sum_{\substack{i=1,3,5,7;\\ \alpha}}f_y^i\int d\tilde k d\tilde k' \ c_{i\alpha}^\dagger (\tilde k) c_{i+1,\alpha}(\tilde k')\phi_y^\dagger (\tilde k'-\tilde k) \nonumber\\
&+g_c\sum_{\substack{i=1,2,3,4;\\ \alpha}}f_x^i\int d\tilde k d\tilde k' \ c_{i\alpha}^\dagger (\tilde k) c_{i+4,\alpha}(\tilde k')\phi_x^\dagger (\tilde k-\tilde k') \nonumber\\
&+h.c.,
\label{cf}
\end{align}
where $c_{i\alpha}$ is a fermion field with $i$ labeling hot spots and $\alpha$ labeling spin projections. Hot spots $i$ and $i+1$ are separated by CDW momentum $Q_y$, and hot spots $i$ and $i+4$ are separated by $Q_x$ (see Fig.\ \ref{fig:4}). The scalar field $\phi_{x,y}$ is a CDW order parameter field with momenta near $Q_{x}/Q_{y}$. In Eq. (\ref{cf}) we have used shorthands $\tilde k=(\omega_m,{\bf \tilde k})$ and $\tilde q=(\Omega_m,{\bf \tilde q})$, where $\omega_m(\Omega_m)$ are fermionic (bosonic) Matsubara frequencies. The bosonic momentum $\bf \tilde  q$ is measured as the deviation from CDW momenta $Q_x$ or $Q_y$, and the fermionic momentum $\bf k$ is measured as the deviation from the corresponding hot spot.  The form-factors $f_{x,y}^i$ determine relative magnitude and sign between CDW orders in different hot regions.
   In general, the CDW order has both $d$-wave and $s$-wave components.
       A pure $d$-wave order would correspond to $f_{y}^{1,5}=-f_y^{3,7}=1$ and $f_x^{1,2}=-f_x^{3,4}=1$.

We assume, like it is done in the spin-fermion model, that static charge susceptibility comes from fermions with energies larger than the one relevant to  superconductivity and approximate it by a simple Ornstein-Zernike form $\chi_{c}=\chi_{0c}/({\tilde q}_x^2+{\tilde q}_y^2+\xi_c^{-2})$, where $\xi_c$ is the CDW correlation length. The prefactors for ${\tilde q}_x^2$ and ${\tilde q}_y^2$ may in general differ~\cite{damascelli_new} because $Q_x$ and $Q_y$ are not along Brillouin zone diagonals, but this difference can be absorbed into rescaling of ${\tilde q}$ and does not affect our analysis.

The coupling $g_c$ is a phenomenological charge-fermion coupling constant.  The corresponding
 charge-mediated 4-fermion interaction term in the Hamiltonian is (we set ${\bf Q} =Q_y$ for definiteness)
 \beq
{\cal H}_{\rm int} = -U^{\rm eff}_c ({\bf q}) \sum_{k,p} c^\dagger_{k, \alpha} c^\dagger_{p,\gamma} c_{k-q, \delta} c_{k+q,\beta} \delta_{\alpha \beta} \delta_{\gamma\delta}
 \label{ch_7}
 \eeq
 with
 \beq
 U^{\rm eff}_c ({\bf q}) =  g^2_c \chi_c ({\bf q}) = 
  \frac{{\bar g}_c}{\xi^{-2}_c + ({\bf q}-{\bf Q}_y)^2}.
 \label{ch_8}
 \eeq
  The effective coupling ${\bar g}_c = g^2_c \chi_{0c}$,  and the sign convention is such that the interaction appears with
   a factor $-1$ in the diagrammatic theory.

   The charge-fermion model is defined self-consistently when its fluctuations cannot modify the physics at
   lattice energies, and this requires that
 $\bar g_c$ must be small compared to the fermionic bandwidth.

\subsection{Normal state analysis}
\begin{figure}[h]
\includegraphics[width=1\columnwidth]{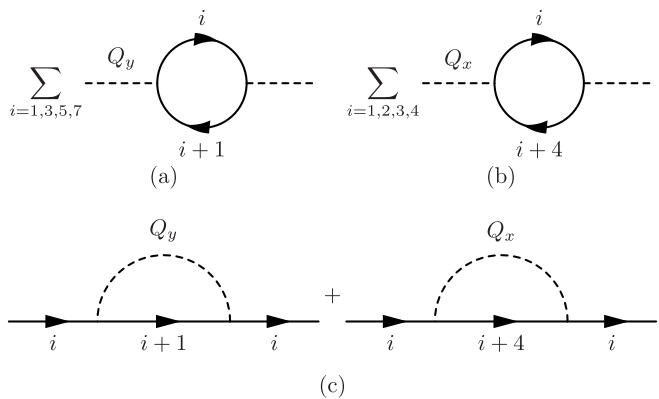}
\caption{The one-loop diagrams for the bosonic self-energy $\Pi_{x,y}$ [Panels (a), (b)] and fermionic self-energy $\Sigma_i$ [Panel (c)].}
\label{fig:5}
\end{figure}
For the computation of superconducting $T^{\rm ch}_c$, mediated by charge fluctuations, we need to know normal state properties.  We use Eq.\ (\ref{cf}) and compute
self-energies for the bosonic fields $\phi_x$ and $\phi_y$ and for the fermionic field
 We show corresponding diagrams in Fig.\ \ref{fig:5}. We compute basonic and fermionic self-energies in a self-consistent fashion, like it was done
   in the earlier works on the spin-fermion model~\cite{acs,ms,2kf}.
   Namely, we
    first evaluate one-loop bosonic self-energy (the bosonic polarization operator)
  using free fermions and show that it has the form of Landau damping, then  use the full dynamical bosonic propagator to calculate one-loop fermionic self-energy and show that it is strong but predominantly depends on frequency,  and then verify that frequency dependent fermionic self-energy does not affect the Landau damping.  This self-consistent procedure becomes exact if we neglect subleading terms in the self-energy, which depend on the fermionic dispersion $\epsilon_k$.  This can be rigorously justified if we extend the model to
   $M$ fermionic
  flavors and take the limit $M \to \infty$
  (e.g. Ref.\ \onlinecite{ms}), or extend the number of pairs of hot spots from $4$ to $N$ and take the limit $N \to \infty$ (e.g. Ref.\ \onlinecite{acs}).  We will use the latter extension to justify our analysis.

\subsubsection{bosonic polarization operators}
The bosonic polarization operator for  $\phi_y$ is given by the diagram in Fig.\ \ref{fig:5}(a), and the expression for $\phi_x$ is related by symmetry.
 The full polarization operator is a sum of contributions from  four pairs of hot fermions which are separated by $Q_y$. These pairs are (1,2), (3,4), (5,6), and (7,8). There are no Umklapp processes for  incommensurate CDW order,  in distinction to SDW case in which ${\bf Q} = (\pi,\pi)$ and Umklapp processes are allowed.

  For the contribution to $\Pi$ from fermion pair (1,2) we obtain
\begin{align}
\Pi_1(\Omega_m,{\tilde {\bf q}})=-2g_c^2T\sum_{\omega_m,\tilde {\bf k}}G_1(\omega_m,{\tilde {\bf k}}) G_2(\omega_m+\Omega_m, {\tilde {\bf k}+\tilde {\bf q}})
\label{pi}
\end{align}
where the factor 2 comes from summation over spin indices.
To simplify the notations we will drop the tildes on momenta hereafter.
 The Green's functions are given by $G_{1,2}=1/(i\omega_m-\epsilon_{1,2})$ and  fermionic dispersions can be written as $\epsilon_{1,2}={\bf v}_{1,2}\cdot{\bf k}$. We transform the momentum integral $dk_x dk_y$ into $d\epsilon_1d\epsilon_2$ by adding the Jacobian $J = 1/|{\bf v}_1\times {\bf v}_2|=1/v_F^2\times (\alpha^2+1)/(2\alpha)$, where $v_F=\sqrt{v_x^2+v_y^2}$. Because this Jacobian is independent of $\bf q$, the polarization operator in this approximation is a function of frequency only. We subtract from $\Pi_1(\Omega_m)$ its frequency-independent part $\Pi_1(0)$, which only renormalizes the position of the CDW QCP and is not of interest to us. The subtraction makes integral over internal
  frequency convergent, and evaluating the integrals we obtain
  at zero temperature
\begin{widetext}\begin{align}
\Pi_{1}(\Omega_m)=&-\frac{g_c^2}{4\pi^3 v_F^2}\frac{\alpha^2+1}{\alpha}\int d\omega_m d\epsilon_1d\epsilon_2\[\frac{1}{(i\omega_m-\epsilon_1)[i(\omega_m+\Omega_m)-\epsilon_2]}-\frac{1}{(i\omega_m-\epsilon_1)(i\omega_m-\epsilon_2)}\] \nonumber\\
&=\frac{g_c^2}{4\pi v_F^2}\frac{\alpha^2+1}{\alpha}\int d\omega_m [\sgn{(\omega_m)}\sgn(\omega_m+\Omega_m)-1] \nonumber\\
&=-\frac{g_c^2}{4\pi v_F^2}\frac{\alpha^2+1}{\alpha}|\Omega_m|.
\label{pi1}
\end{align}
\end{widetext}
This is a conventional Landau damping term. We do the same calculation for hot spot pairs (3,4), (5,6), and (7,8). Because the Jacobians $1/|{\bf v}_1\times {\bf v}_2|=1/|{\bf v}_3\times {\bf v}_4|=1/|{\bf v}_5\times {\bf v}_6|=1/|{\bf v}_7\times {\bf v}_8|=1/v_F^2\times(\alpha^2+1)/(2\alpha)$,  all contributions
 are the same as (\ref{pi1}). Therefore $\Pi_y(\Omega_m)=4 \Pi_1(\Omega_m)$. It is easy to verify that for $\phi_x$ the self-energy is the same as for $\phi_y$, i.e., $\Pi_x=\Pi_y$.

 Including the polarization operator into the propagators for $\phi_{x}$ and $\phi_y$ fields, we obtain
\begin{align}
\chi_c(\Omega_m,{\bf q})=\frac{\chi_{0c}}{\xi^{-2}_c + q_x^2+q_y^2+\gamma_c|\Omega_m|},
\label{chic}
\end{align}
where
\begin{align}
\gamma_c=\frac{4\bar g_c}{4\pi v_F^2}\frac{\alpha^2+1}{\alpha}
\label{gamma}
\end{align}
and we recall that $\bar g_c=g_c^2\chi_{0c}$. The overall factor of 4 in the numerator of (\ref{gamma}) is the number of pairs of hot fermions.  To extend the model to large $N$ one just has to replace $4$ by $N$.  We will use this extension below.

The total dynamical interaction mediated by charge fluctuations can then be written as
 \beq
 U^{\rm eff}_c ({\bf q},\Omega_m) =  g^2_c \chi ({\bf q}) =  \frac{{\bar g}_c}{\xi^{-2}_c + ({\bf q}-{\bf Q}_x)^2 + \gamma_c |\Omega_m|}.
 \label{ch_8_1}
 \eeq

\subsubsection{fermionic self-energy}
 We now use Eq.\ (\ref{chic}) and evaluate one-loop fermionic self-energy. The corresponding diagram is presented in Fig.\ \ref{fig:5}(c). Observe that for any
   hot fermion,  interactions mediated by both bosonic fields, $\phi_x$ and $\phi_y$, contribute to the self-energy. For example, for a fermion at hot spot 1, $\phi_x$ and $\phi_y$ scatter it to hot spots 2 and 5, respectively.

The self-energy depends on the location of a fermion on the FS and on the distance to CDW QCP.  Below we will be interested in superconductivity right at CDW QCP, hence we will need the self-energy right at this point.  Accordingly,
 we  set $\xi^{-1}_c =0$ in the charge fluctuation propagator.

For the self-energy contribution to a fermion at hot spot 1 from $Q_y$ scattering, we have
\begin{align}
\Sigma_y({\bf k},\omega_m)= T\sum_{\omega_m',{\bf k'}}U^{\rm eff}_c({\bf k-k'}, \omega_m-\omega_m')G_2(\omega_m',{\bf k'}),
\label{sigma}
\end{align}
where $\bf k$ is the deviation from hot spot 1.
We
 place $\bf k$ on the FS, i.e., set $k_{\perp}\equiv \hat {\bf v}_1\cdot {\bf k}=0$, which gives $k_y=-\alpha k_x$. At $T=0$, we rewrite Eq.\ (\ref{sigma}) as
\begin{align}
 &\Sigma_y(k_{\|},\omega_m)=\frac{-\bar g_c}{8\pi^3}\int\frac{d\omega_m' dk'_{\perp} dk'_{\|}}{i\omega_m'-v_Fk'_{\perp}}\nonumber\\
 &\times\frac{1}{({ {\bar k}_{\|}}-{k'_{\|}})^2+({ {\bar k}_{\perp}}-{k'_{\perp}})^2+\gamma_c|\omega_m-\omega_m'|},
 \label{eq10}
 \end{align}
where  $k'_{\perp}$ and $k'_{\|}$ are perpendicular and parallel components of $\bf k'$ with respect to the Fermi surface at hot spot 2, i.e.,  $k'_{\perp}\equiv \hat {\bf v}_2\cdot {\bf k}'$, and ${\bar k}_{\|}$ and ${\bar k}_{\perp}$ are components of external ${\bf k}$,
defined relative to the FS at the hot spot 2, i.e.,
 ${\bar k}_{\perp}\equiv \hat {\bf v}_2\cdot {\bf k}$, where ${\bf v}_2=(v_x,-v_y)$.  Using $k_y=-\alpha k_x$
  we obtain
\begin{align}
{\bar k}_{\perp}=2\alpha k_{\|}/(\alpha^2+1).
\label{yw_new}
\end{align}

We integrate over $k'_{\perp}$ first, and complete the integration contour over the half plane with only one pole. We obtain
\begin{align}
&\Sigma_y(k_{\|},\omega_m)=\frac{i \bar g_c }{8\pi^2 v_F}\int \frac{ dk'_{\|}d\omega_m' \sgn(\omega_m')}{\sqrt{({ {\bar k}_{\|}}-{k'_{\|}})^2+\gamma_c|\omega_m-\omega_m'|}}\nonumber\\
&\frac{1}{\sqrt{({ {\bar k}_{\|}}-{k'_{\|}})^2+\gamma_c|\omega_m-\omega_m'|}+[|\omega_m'|/v_F+i{\bar k}_{\perp}\sgn(\omega_m')]}
\label{sigma2}
\end{align}
 We will see that typical internal frequencies $\omega'_{m}$ are of the same order as external $\omega_m$. Then, at small enough frequencies one
can safely neglect the $|\omega_m'|/v_F$ term in the denominator.
With this simplification we obtain
\begin{align}
&\Sigma_y(k_{\|},\omega_m)\nonumber\\
&=\frac{i \bar g_c }{8\pi^2 v_F}\int \frac{dk'_{\|}d\omega_m' \sgn(\omega_m')}{({\bar k}_{\|}-k'_{\|})^2+{\gamma_c |\omega_m-\omega_m'|}+{\bar k}^2_{\perp}}\nonumber\\
&=\frac{i \bar g_c }{2\pi v_F\gamma_c}\sgn(\omega_m)\[\sqrt{\gamma_c|\omega_m|+{\bar k}^2_{\perp}}-|{\bar k}_{\perp}|\].
\label{sigmay}
\end{align}
Plugging Eq.\ (\ref{yw_new}) into Eq.\ (\ref{sigmay}) we finally obtain
\begin{align}
\Sigma_y(k_{\|},\omega_m)=&\frac{i \bar g_c }{2\pi v_F\gamma_c}\sgn(\omega_m)\nonumber\\
&\times\[\sqrt{\gamma_c|\omega_m|+\(\frac{2\alpha k_\|}{\alpha^2+1}\)^2}-\left|\frac{2\alpha k_\|}{\alpha^2+1}\right|\]\nonumber\\
=& \frac{2 i v_F}{N} \frac{\alpha}{\alpha^2 +1} \sgn(\omega_m)\nonumber\\
&\times\[\sqrt{\gamma_c|\omega_m|+\(\frac{2\alpha k_\|}{\alpha^2+1}\)^2}-\left|\frac{2\alpha k_\|}{\alpha^2+1}\right|\].
\end{align}

The self-energy from $Q_x$ scattering is obtained in the same way:
\begin{align}
\Sigma_x({\bf k},\omega_m)=-\bar g_cT\sum_{\omega_m',{\bf k'}}\chi_c(\omega_m-\omega_m',{\bf k-k'})G_5(\omega_m',{\bf k'}).
\label{sigmax}
\end{align}
 As Fermi velocities at hot spot 5 and 2 are antiparallel, we have $G_5(\omega_m',{\bf k'})=G_2(\omega_m',-{\bf k'})$. Comparing Eqs.\ (\ref{sigma}) and (\ref{sigmax}), we then immediately find that $\Sigma_x=\Sigma_y$. Combining the two we obtain
\begin{align}
\Sigma(k_{\|},\omega_m)=&\frac{i \bar g_c }{\pi v_F\gamma_c}\sgn(\omega_m)\nonumber\\
&\times\[\sqrt{\gamma_c|\omega_m|+\(\frac{2\alpha k_\|}{\alpha^2+1}\)^2}-\left|\frac{2\alpha k_\|}{\alpha^2+1}\right|\]\nonumber\\
=& \frac{4 i v_F}{N} \frac{\alpha}{\alpha^2 +1} \sgn(\omega_m)\nonumber\\
&\times\[\sqrt{\gamma_c|\omega_m|+\(\frac{2\alpha k_\|}{\alpha^2+1}\)^2}-\left|\frac{2\alpha k_\|}{\alpha^2+1}\right|\].
\label{sigmaxy}
\end{align}
It is easy to verify that this results holds for around all hot regions 1-8, and in each region $k_{\|}$ is the deviation from the corresponding hot spot along the FS.
The functional form of the self-energy as in Eq.\ (\ref{sigmaxy}) was first obtained for the spin-fermion model in Ref.\ \onlinecite{acs} (for $\alpha =1$) and Ref.\ \onlinecite{ms} (for arbitrary $\alpha$).   Right at a hot spot, the fermionic self-energy has a non-Fermi liquid (NFL) form:
\begin{align}
\Sigma(0,\omega_m)=i\sgn(\omega_m)\sqrt{\omega_{0c}|\omega_m|},
\end{align}
where
$\omega_{0c}=(4/N) \bar g_c/\pi\times\alpha/(\alpha^2+1)$. Away from a hot spot, at $k_\|^2\gg\gamma_c|\omega_m|$, the self-energy $\Sigma(\omega_m,k_\|)$ retains a Fermi liquid (FL) form at the smallest $\omega_m$, i.e., we have
\begin{align}
\Sigma(\omega_m,k_{\|})=&\frac{i\omega_m}{|k_\||}\(\frac{\bar g_c}{\pi v_F}\frac{\alpha^2+1}{4\alpha}\) + O(\omega^2_m)
\label{sigmaxy0}
\end{align}
One can easily verify that the inclusion of fermionic self-energy $\Sigma(k,\omega_m)\propto i \sgn (\omega_m)$ will not change the polarization operator, 
i.e., $\Pi (\Omega_m)$ retains the same form even if we compute it using dressed fermions.

To verify self-consistency of the calculations, we also computed the self-energy away from the FS. We found $\Sigma (k_\perp,0) \propto (1/N) v_F k_\perp \log({\Lambda/|v_F k_\perp|})$, where $\Lambda$ is the upper cutoff in momentum integration.  The presence of the logarithm implies that Fermi velocity also acquires singular renormalization at CDW QCP (Refs.\ \onlinecite{acs,ms}).  This singularity breaks self-consistency of one-loop calculation of fermionic and bosonic self-energies if we keep $N$ finite, but self-consistent procedure still remains rigorously justified at this loop order if we set $N \to \infty$.
The situation gets more complex at higher loop orders due to special role of forward scattering and backscattering processes which give rise to the dependence of $\Sigma$ on $k_{\perp}$ without the factor $1/N$ (Refs.\ \onlinecite{sslee,ms_1,senthil,garst}).
  How important are these effects in unclear and in this work we restrict with one-loop self-energy.

\subsection{The pairing problem}
We now use the normal state results as inputs for the analysis of the pairing  mediated by CDW fluctuations with momenta around $Q_x$ and $Q_y$.
To leading order in $1/N$, the pairing problem can be analyzed without vertex corrections,  by summing up the ladder series of diagrams in the particle-particle channel~\cite{acs,ms}.

\begin{figure}[h]
\includegraphics[width=\columnwidth]{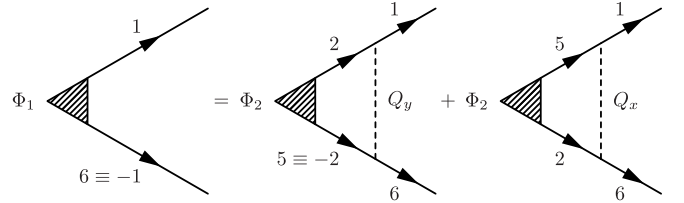}
\caption{Diagrammatic representation of the coupled ladder equations for superconducting order parameters $\Phi_1$ and $\Phi_2$, which involve fermions in hot regions 1 and 2 in Fig. 1.}
\label{fig:sc}
\end{figure}

We first focus on the momentum region where hot spots 1,2,
and $5 = -2$, and $6=-1$ are located (see Fig.\ \ref{fig:4}).
  We introduce  superconducting order parameters $\Phi_1\sim\langle c_1 c_6\rangle = \langle c_1 c_{-1}\rangle$ and $\Phi_2\sim\langle c_2 c_5\rangle =
  \langle c_2 c_{-2}\rangle$ and obtain in the standard way a set of coupled gap equations for the two condensates.
  The interactions with momentum transfer $Q_x=(Q,0)$ and $Q_y=(0,Q)$ connect hot spots  1,5 with 2,6, and hot spots 1,2 with 5,6, correspondingly.  As the consequence, the interactions relate $\Phi_1$ with $\Phi_2$ and vise versa. We show the equation for $\Phi_1$  diagrammatically in Fig.\ \ref{fig:sc}. In analytical form, we have
\begin{widetext}\begin{align}
\Phi_1(k)=
&T\sum_{k'}U^{\rm eff}_c (k-k')\times[G_2(k')G_5(-k')\Phi_2(k')+G_2(-k')G_5(k')\Phi_2(-k')]
\label{sc1}
\end{align}
\end{widetext}
where $k=(\omega_m,{\bf k})$ and $k'=(\omega_m',{\bf k'})$ and ${\bf k},{\bf k}'$ are momentum deviations from the corresponding hot spots. The fermionic Green's function are given by $G_i(k)=
1/[i\omega_m-\epsilon_i({\bf k})+\Sigma_i(k)]$. The equation for $\Phi_2$ in terms of $\Phi_1$ has the same form, and thus $\Phi_1$ and $\Phi_2$ have the same magnitude.

 Because the two kernels in the Eq.\ (\ref{sc1}) (the prefactors for $\Phi_2$ in the right hand side) are both positive (we recall that $U_{\rm eff}$ is 
  positive),
   the
    $U(1)$ order parameters
  $\Phi_1$ and $\Phi_2$ have the same phase, i.e.,  $\Phi_1 =\Phi_2$.
    By the same token, the SC order parameters in the momentum range near hot spots 3,4,7,8, namely $\Phi_3\sim\langle c_3 c_8\rangle
  =\langle c_3 c_{-3}\rangle$ and $\Phi_4 \sim \langle c_4 c_7\rangle =\langle c_4 c_{-4}\rangle$ are also equal.  The kernels of the gap equations in the regions 1,2,5,6 and 3,4,7,8 are the same, hence the magnitudes of $\Phi_1 = \Phi_2$ and $\Phi_3=\Phi_4$ are identical.  However, there is no specification of the relative phase between superconducting order parameters in the two regions.
  Setting aside more exotic possibilities of phase difference equal to a fraction of $\pi$, we are left with two options for the pairing symmetry: an $s$-wave,
    for which the phases of $\Phi_1$ and $\Phi_3$ are identical, and a
    $d_{x^2-y^2}$, for which $\Phi_3 = - \Phi_1$  (see Fig.\ \ref{fig:4}).
    When only CDW-mediated interaction is considered, the two pairing states are degenerate.
   This has been noticed before~\cite{italians}, and it was argued that the degeneracy is lifted by other interactions, e.g.,  antiferromagnetic spin fluctuations would favor $d$-wave.

We now proceed with the calculation of $T^{\rm ch}_c$.  We assume and then verify that the $\Phi_1(k) =\Phi_2(k) = \Phi (k)$ are even functions of momentum $\bf k$. The linearized gap equation (\ref{sc1}), whose solution exists right at $T=T_c$,  then becomes
\begin{align}
\Phi(k)=&2T\sum_{k'}|U^{\rm eff}_c (k-k')|G_2(k')G_5(-k')\Phi(k').
\label{sc2}
\end{align}
In the right hand side of Eq.\ (\ref{sc2}) we first integrate over momentum transverse to the FS. Neglecting terms small in $1/N$, we obtain
\begin{align}
\Phi(\omega_m, k_\|)=&\frac{\bar g_c T}{v_F}\sum_{m'}\int \frac{dk'_{\|}}{2\pi}\frac{\Phi(\omega_m',k'_\|)}{|\omega_m'-i\Sigma(\omega_m', k'_{\|})|}\nonumber\\
&\times\frac{1}{k_\|^2+k_\|^{\prime2}-2\beta k_\|k_\|'+\gamma_c|\omega_m-\omega_m'|},
\label{sc3}
\end{align}
where  $\beta=(1-\alpha^2)/(1+\alpha^2)$. This factor appears in the last term in (\ref{sc3}) because
$k_\|$ and $k'_\|$ are parallel components of momenta
 in {\it different} segments of the FS, namely
 near hot spots 1 and 2, respectively.

A similar gap equation has been analyzed in the context of spin-mediated pairing near SDW QCP~\cite{acs,ms,wang}.  To make this paper self-contained, we present some details of the computation of $T^{\rm ch}_c$ in our case.

It is usually more convenient not to  solve Eq.\ (\ref{sc3}) directly, but to add to the right hand side of the gap equation an infinitesimal pairing condensate $\Phi_0$ and compute pairing susceptibility $\chi_{pp} = \Phi/\Phi_0$.  The transition temperature $T_c$ is the one at which pairing susceptibility diverges.  This approach has an advantage in that the pairing susceptibility can be analyzed within perturbation theory.

 The first iteration gives
\begin{align}
  \Phi(\omega_m\sim T,0)=\Phi_0\(1+\frac{l}{2\pi}\log^2\frac{\omega_{0c}}{T}\),~~l=\frac{2\alpha}{\alpha^2+1},
  \label{sc4}
  \end{align}
 where, we remind, $\omega_{0c} \sim {\bar g}_c
 $ is the upper edge of NFL behavior.
   We note that neither the coupling constant $\bar g_c$ nor $1/N$
   directly appear in Eq. (\ref{sc4}),
    i.e., once temperature is expressed in units of $\omega_{0c}$, the renormalization of $\Phi_0$ is fully universal.
  The presence of $\log^2$ term (i.e., one extra power of $\log$ compared to BCS theory) is the consequence of the singular dependence of the fermionic self-energy on momentum along the FS. The $\log^2$ term comes from  momentum range where $\Sigma(\omega_m,k_\|)\sim i\omega_m/|k_\||$ [see Eq.\ (\ref{sigmaxy0})]. At
   $\bar g_c /v_F \gg |k_\||\gg\sqrt{\gamma_c|\omega_{0c}|}$
    the term $1/|\omega_m'-i\Sigma(\omega_m', k'_{\|})|$ in Eq.\ (\ref{sc3}) scales as $|k_\|/\omega_m|$. To logarithmical accuracy, the momentum integral over $k_\|'$ in Eq.\ (\ref{sc3}) yields $\int_{\gamma_c |\omega_m'|} dk_\|^2/k_\|^2\propto \log |\omega_m'|$, and the frequency integral over $\omega_m'$ then yields $\int_T (\log|\omega_m'|)/|\omega_m'| d\omega_m'\propto \log^2 (1/T)$.  For the spin-fermion model, this result was first obtained in Ref.\ \onlinecite{ms}.

  The $\log^2 T$ renormalization of the pairing vertex has been found in other contexts~\cite{color,joerg,raghu, d+id}. To see how it is relevant for $T_c$  one has to go beyond one loop order.   To get the insight, we first
  consider  the ``weak coupling" limit by formally replacing the actual coupling $l=2\alpha/(\alpha^2+1)$  by an effective $l_\epsilon=2\epsilon\alpha/(\alpha^2+1)$ and taking the limit
  $\epsilon\ll 1$. In this limit,  series of $\log^2$ renormalizations can be summed up explicitly, and the result is
  $\Phi=\Phi_0 e^{l_\epsilon/(2\pi) \log^2{(\omega_{0c}/T)}}$. We see that, at the $\log^2$ level,  the pairing susceptibility does increase with decreasing $T$, but it does not diverges at any finite $T$. One then has to go beyond the $\log^2$  approximation and include subleading $O(\log)$ terms.
  In the weak coupling limit $\epsilon\ll 1$ this can be done rigorously, along the lines specified in Ref.\ \onlinecite{wang}, and the result is that subleading $O(\log)$ terms do give rise to the divergence of the pairing susceptibility at a finite $T^{\rm ch}_c$ given by
 \begin{align}
T^{\rm ch}_c\sim \omega_{0c} e^{-1/\epsilon}.
\label{ch_1}
\end{align}
The exponential dependence is the same as in  BCS formula, which is not accidental because in the limit $\epsilon\ll 1$ the main contribution to superconductivity comes from fermions away from the hot regions, where self-energy has FL form.
However, in distinction to  BCS formula, the prefactor  $\omega_{0c}$ is not the upper cutoff for the attraction but rather the scale
  set by the coupling constant $\bar g_c$. The proportionality of  $T^{\rm ch}_c$ to the coupling $\bar g_c$ is the fingerprint of the pairing near a quantum-critical point~\cite{acf, max_new}.

For the physical case $\epsilon=1$, we expect from (\ref{ch_1}) $T^{\rm ch}_c \sim \omega_{0c}$. To obtain the exact relation
we solved Eq.\ (\ref{sc3}) directly, using the finite-temperature form of the fermionic self-energy. Typical internal momenta and typical internal frequency are of order, $k_\|^2\sim \gamma_c \omega_m\sim \gamma \omega_{0c}$ and $\omega \sim \omega_{0c}$.  In this situation,  fermions from both NFL and FL regions contribute to the pairing. For numerical evaluation of $T^{\rm ch}_c$ we extracted the fermionic dispersion in hot regions from ARPES data for Bi$_2$Sr$_2$CaCu$_2$O$_{8+x}$ (Ref.\ \onlinecite{mike_tb}) and obtained $\alpha=0.074$. Using this value for $\alpha$ we obtained numerically
\begin{align}
T^{\rm ch}_c=0.0025 \bar g_c.
\end{align}
 For comparison, in the spin-fermion model the critical temperature at SDW QCP is~\cite{wang} $T^{\rm sp}_c=0.0073 \bar g_s$ ($T^{\rm sp}_c \sim 140~{\rm K}$ for ${\bar g}_s \sim 1.7~{\rm eV}$). We see that  $T_c$  in spin-fermion and  charge-fermion models on top of the corresponding QCP are  comparable if ${\bar g}_c \geq {\bar g}_s$. If this is the case,
  then CDW fluctuations give rise to substantial enhancement of the superconducting $T_c$ around CDW QCP.
  One also should keep in mind that the result we quoted for $T^{\rm sp}_c$ due to spin fluctuation exchange is $T^{\rm sp}_c$ right on top of SDW QCP.
   Near the CDW QCP, magnetic $\xi_s$ is finite and spin-mediated $T^{\rm sp}_c$ is reduced.

Away from the CDW QCP,  charge-fluctuation exchange preserves FL behavior and charge-mediated $T_c$ drops and eventually  follows weak coupling BCS formula.  The self-energy at a finite charge correlation length $\xi_c$ is modified compared to Eq. (\ref{sigmaxy}) and is given by
\begin{widetext}
\begin{align}
\Sigma(k_{\|},\omega_m)=&\frac{i \bar g_c }{\pi v_F\gamma_c}\sgn(\omega_m)\[\sqrt{\gamma_c|\omega_m|+\(\frac{2\alpha k_\|}{\alpha^2+1}\)^2 + \xi^{-2}_c}-\sqrt{\left(\frac{2\alpha k_\|}{\alpha^2+1}\right)^2 + \xi^{-2}_c}\].
\label{sigmaxy_1}
\end{align}
\end{widetext}
Now even right at a hot spot (at $k_{\parallel}=0$) the self-energy has a FL form
\beq
\Sigma (\omega_m,0) = \lambda_c (i\omega_m) -i \frac{\omega^2_m}{4\omega_{\rm cf}},
\label{ch_n1}
\eeq
 where
 \beq
 \lambda_c = \frac{{\bar g}_c\xi_c}{2\pi v_F },~~~\text {and}~~~\omega_{\rm cf} =\frac {\xi^{-2}_c}{\gamma_c} = \frac{{\bar g}_c}{4\pi \lambda_c^2} \frac{\alpha}{\alpha^2+1}.
 \label{ch_n2}
 \eeq
 The dimensionless charge fermion coupling $\lambda_c$ (the ratio of ${\bar g}_c$ to typical fermionic energy $v_F \xi^{-1}_c$) decreases
 when $\xi_c$ decreases. Once $\lambda_c \leq 1$, charge-relaxation scale $\omega_{\rm cf}$ becomes the upper energy cutoff for the pairing, and
 $T_c$ follows BCS-Eliashberg-McMillan formula~\cite{mcmillan}
 \beq
 T_c \sim \omega_{\rm cf} ~e^{-\frac{1+\lambda_c}{\lambda_c}}.
 \label{ch_n3}
 \eeq
We present
the numerical result for the
 behavior of  $T^{\rm ch}_c$   as a function of $\xi_c$ in Fig.\ \ref{H} (red line).  A similar reduction of charge-mediated $T^{\rm ch}_c$ is expected on the other side of CDW QCP, in the charge-ordered state.

\section{charge-fermion coupling constant from the spin-fermion model}

  \begin{figure}
\includegraphics[width=\columnwidth]{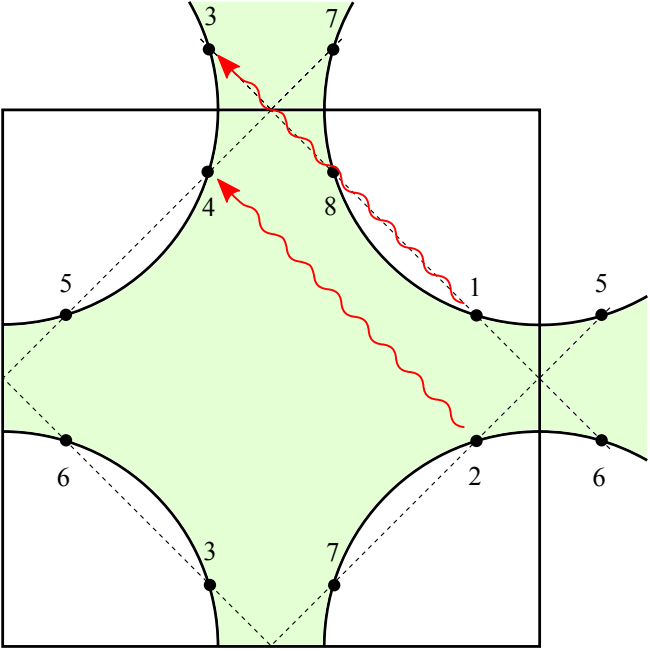}
\caption{Schematic representation of spin-mediated interaction.
 Near an
 antiferromagnetic quantum-critical point in a metal hot fermions scatter into each other by exchanging soft antiferromagnetic spin fluctuations with momentum $(\pi,\pi)$ (the wavy lines).}
\label{fig:0}
\end{figure}

To compare the magnitudes of ${\bar g}_c$ and ${\bar g}_s$ we compute their ratio within a particular microscopic model for charge order in the cuprates. Namely, we assume, as in earlier works by several groups including us~\cite{acs,efetov,acf,acn,ms,haslinger,berg,peter}
 that spin fluctations develop at higher energies than CDW (and superconducting) fluctuations, and CDW order emerges due to spin-fluctuation exchange, as a composite order.

 To this end, we consider a 2D itinerant electron system
  in which the primary interaction between fermions is mediated by soft collective spin fluctuations at the antiferromagnetic momentum ${\bf Q}_\pi=(\pi,\pi)$. Such an interaction,  shown as the wavy line in Fig.\ \ref{fig:0}, scatters fermions between hot spots 2 and 4, 1 and 3, etc, and is proportional to the dynamical spin susceptibility:
\beq
 {\cal H}_{\rm eff} = -U^{\rm eff}_s ({\bf q},\Omega_m) \sum_{k,p} c^\dagger_{k, \alpha} {\vec \sigma}_{\alpha \beta} c_{k+q,\beta} c^\dagger_{p,\gamma} {\vec \sigma}_{\gamma\delta} c_{p-q, \delta}
 \label{ch_2_1}
 \eeq
 where
 \begin{align}
 U^{\rm eff}_s ({\bf q},\Omega_m) =&  g^2_s \chi_s ({\bf q},\Omega_m)\nonumber\\
  =& \frac{{\bar g}_s}{\xi^{-2}_s + ({\bf q}-{\bf Q}_\pi)^2 + \gamma_s|\Omega_m|},
\label{chis}
 \end{align}
where $\gamma_s$ is the corresponding Landau damping coefficient and the scale
${\bar g}_s$ sets the magnitude of $T^{\rm sp}_c$ for spin-mediated
 superconductivity.

Except for special cases, there is no rigorously justified procedure to obtain
 Eq. (\ref{ch_2_1})
  starting from a model of fermions interacting with some short-range
 interaction $U(r)$ because the  main contribution to static part of $U^{\rm eff}_s$ comes from fermions with high energies.

One commonly used approach is to treat static part of spin-mediated interaction phenomenologically and just postulate Ornstein-Zernike form of the static $\chi_s ({\bf q},0)$ (this is the same procedure that we used in the charge-fermion model).
     Once the model with static spin-mediated interaction is established, one can compute the
       dynamical part of  $\chi_s ({\bf q},\Omega_m)$ (the Landau damping term) within this model, as it
       comes from fermions with small energies.  Within this approach, one cannot relate ${\bar g}_s$ with $U(r)$, but one can express
 Landau damping coefficient $\gamma_s$ via ${\bar g}_s$. The relation is~\cite{acs}  $\gamma_s=2 {\bar g}_s [2/(\pi v_F^2)]$, where for comparison with RPA below we
  pulled out factor of 2 due to spin summation.

A complementary approach is to
  treat $U^{\rm eff}_s$ as the charge component of the fully renormalized vertex function $\Gamma_{\alpha\gamma,\beta\delta}({\bf q},\Omega_m)$  at momentum transfer  ${\bf q}$ near ${\bf Q}_\pi$. The vertex function $\Gamma$  is
the opposite of  physical anti-symmetrized interaction
(a direct interaction minus the one with outgoing fermions interchanged).
The vertex function can be obtained in RPA by summing up particular ladder and bubble diagrams which form geometrical series (for details see Ref.\ \onlinecite{repulsion}).  The approach is best understood when $U (r)$ is approximated as on-site Hubbard interaction $U$.  The RPA gives
 \begin{widetext}
  \begin{align}
\Gamma_{\alpha\gamma,\beta\delta}({\bf q},\Omega_m)
  & =
  -
  \frac{U}{1- U^2 \Pi^2 ({\bf q},\Omega_m)} \delta_{\alpha \beta} \delta_{\gamma \delta}
  +
   \frac{U}{1- U \Pi ({\bf q},\Omega_m)} \delta_{\alpha \delta} \delta_{\beta\gamma} \nonumber \\
&=
-
 \frac{U}{2(1 + U \Pi ({\bf q},\Omega_m))}
\delta_{\alpha \beta} \delta_{\gamma\delta}
+\frac{U}{2(1- U \Pi ({\bf q},\Omega_m))} {\vec \sigma}_{\alpha \beta}\cdot {\vec \sigma}_{\gamma\delta}
\label{ch_2}
\end{align}
\end{widetext}
where to split the vertex into spin and charge parts we used
${\vec \sigma}_{\alpha\beta}\cdot {\vec \sigma}_{\gamma\delta} = - \delta_{\alpha\beta} \delta_{\gamma\delta} +2 \delta_{\alpha\delta}\delta_{\beta\gamma}$
For  a repulsive interaction $U >0$, and the interaction in the spin channel is enhanced and at large enough $U$
diverges at ${\bf q}$, at which static $\Pi ({\bf q},0)>0$ is at maximum.  We assume that the maximum of $\Pi ({\bf q},0)$ is at ${\bf q} = {\bf Q}_\pi$.

Near $\Pi ({\bf Q}_\pi,0) = 1/U$, the interaction in the spin channel well exceeds the one in the charge channel, and one can keep only the spin component of the interaction, i.e., approximate the dressed interaction by Eq.
 (\ref{ch_2_1}) with
\beq
U_s^{\rm eff}{\vec \sigma}_{\alpha \beta} \cdot{\vec \sigma}_{\gamma\delta}=\frac{U}{2(1- U \Pi ({\bf q},\Omega_m))} {\vec \sigma}_{\alpha \beta} \cdot{\vec \sigma}_{\gamma\delta}.
 \label{ch_2_2}
 \eeq
Expanding the polarization operator near antiferromagnetic momentum and zero frequency, we obtain
\beq
\Pi ({\bf q}, \Omega_m) = \Pi ({\bf Q}_\pi, 0) - C_\pi ({\bf q} - {\bf Q}_\pi)^2 - \frac{2 |\Omega_m|}{\pi v^2_F}
\label{ch_3}
\eeq
(the last term comes from fermions near the FS and the prefactor for $\Omega_m$ term is known exactly).
Substituting this form into  (\ref{ch_2_2})  we obtain after simple manipulations
 the same 
 $U_s^{\rm eff} ({\bf q},\Omega_m)$ as in Eq. (\ref{chis}) with
 \beq
 {\bar g}_s = \frac{1}{2 C_\pi},~~ \xi_s^{-2} = \frac{1-U \Pi ({\bf Q}_\pi,0)}{U C_\pi},~~\gamma_s = 2 {\bar g}_s  \frac{2}{\pi v^2_F}
  \label{ch_4}
  \eeq
  We see that the expression for the Landau damping coefficient is exactly the same as in the other (semi-phenomenological) approach, the only new element of RPA is that ${\bar g}_s =1/(2 C_\pi)$ is related to the behavior of the static polarization bubble.   Formally, ${\bar g}_s$  doesn't depend on $U$, but in reality $C_\pi$ is of the same order as $\Pi ({\bf Q}_\pi,0)$ (in units where lattice constant $a=1$), and the latter is approximately $1/U$ near a SDW QCP. As a result, ${\bar g}_s$ is fact is of order $U$.

  The interaction mediated by spin fluctuations gives rise to $d$-wave superconductivity and to fermionic self-energy. In the FL regime, $\Sigma (\omega_m) \approx \lambda_s \omega_m$, where $\lambda_s =  3 {\bar g}/(4\pi v_F \xi^{-1}_s)$. This self-energy gives rise to mass renormalization $m^*/m = 1 + \lambda_s$ and to quasiparticle residue $1/Z = 1/(1+ \lambda_s)$. We will include this renormalization into the calculations below.

   We now proceed to construct the interaction in the CDW channel.

The CDW instability with the ordering momentum $Q_x = (Q,0)$ and $Q_y =(0,Q)$ emerges in this approach as a preliminary collective instability
  at a finite $\xi_s$, due to spin-fluctuation exchange.
    A way to obtain CDW instability is to introduce infinitesimal CDW field $\Delta_k^Q$, which couples to incommensurate component of charge density as
     $\Delta^Q_k c^\dagger_{k-Q/2,\alpha} \delta_{\alpha\beta}
c_{k+Q/2,\beta}$, and compute susceptibility with respect to this field.  This has been done~\cite{laplaca,charge,debanjan} by summing up ladder series of  renormalizations due to spin-fluctuation exchange.  Each act of spin-fluctuation exchange transforms hot fermions near, say, hot points 1 and 2 in Fig.\ \ref{fig:0}
into another set of hot fermions near the points 3 and 4.  The set 3,4 is generally different from the set 1,2 (because directions of the Fermi velocities are different), so to obtain the susceptibility one has to solve the set of two coupled equations for fully renormalized $\Delta^Q_k$ with the center of mass momentum ${\bf k}$ either between points 1 and 2 or between points 3 and 4.

There is no rigorous justification why one should restrict with only ladder diagrams, even at large $N$. The first
 non-ladder diagram is small numerically, but not parametrically, compared to the ladder diagram of the same loop order.  Accordingly, there is no point to keep $N$ as artificially large parameter, and  in this section we set the number of  pairs of hot spots $N$ to their actual value $N=4$.

\begin{figure}
 \includegraphics[width=0.8\columnwidth]{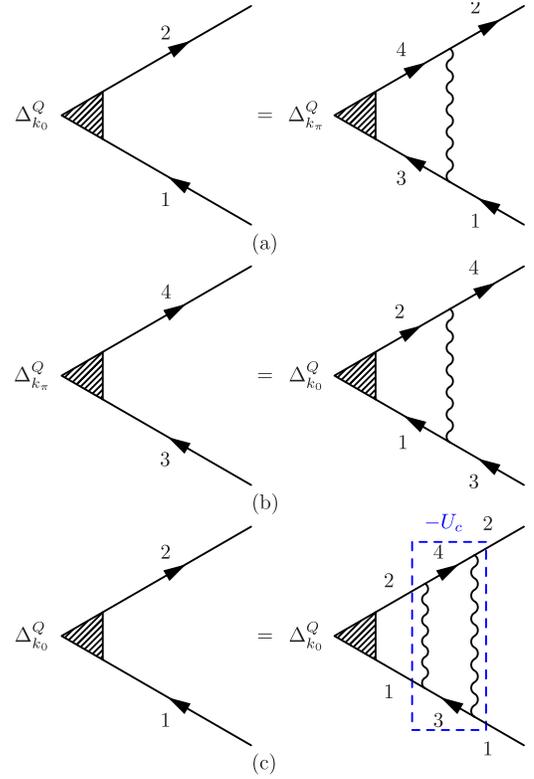}
 \caption{The linearized ``gap" equation for the CDW order parameter $\Delta_k^Q\sim\langle c^\dagger_{k+Q/2} c_{k-Q/2}\rangle$. We  define  the center-of-mass momentum of hot spots 1 and 2 as $k_0$ and  that of hot spots 3 and 4 as $k_\pi$. Panels (a) and (b): the coupled
  gap equations for $\Delta_{k_0}^Q$ and $\Delta_{k_\pi}^Q$. Panel (c): The  gap equation for $\Delta_{k_0}^Q$ only, obtained by combining  panels (a) and (b). We treat the composite object in the dashed frame as the effective interaction $U_c$ (the prefactor $-1$ reflects that interaction
   appears in the diagram with the minus sign).}
 \label{fig:cdw}
 \end{figure}

We present the diagrammatic representation of this set of equations in Fig.\ \ref{fig:cdw}(a,b). The linear ``gap" equation for $\Delta_k^Q$ has been analyzed in Ref.\ \onlinecite{charge} and the outcome is that the CDW susceptibility diverges at a finite $T_{\rm cdw}$ before the system develops CDW order.  The critical temperature $T_{\rm cdw}$ decreases as $\xi_s$ decreases
 and vanishes at a finite critical $\xi_s$, setting up a CDW QCP at some distance away from SDW QCP~\cite{charge_new}
 (see Fig.\ \ref{phases_cdw}).

\begin{figure}
\includegraphics[width=0.5\columnwidth]{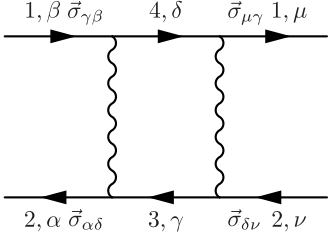}
\caption{The diagrammatic representation of the composite effective interaction $U_c$ (same as in the dashed frame in Fig. \ref{fig:cdw}.
This effective interaction
 is a convolution of a particle-hole bubble and two antiferromagnetic spin-fluctuation propagators.}
\label{fig:x}
\end{figure}

Alternatively, one can combine pairs of subsequent renormalizations of $\Delta^Q_k$ into new effective interaction at {\it small} momentum transfer
[see Fig.\ \ref{fig:cdw}(c)], which we label as $U_c$ and  show graphically in Fig.\ \ref{fig:x}.
 This composite effective interaction is the convolution of two fermionic propagators and two spin-fluctuation propagators.
Explicit calculation shows~\cite{charge_new,maslov} that
 $U_c$ is
  numerically smaller but parameter-wise of the same order as a single spin-fluctuation propagator -- one extra power of ${\bar g}_s$ in the numerator gets cancelled out by the Landau damping coefficient $\gamma_s \propto {\bar g}_s$ in the denominator.
This effective interaction is repulsive ($U_c >0$) because the
  polarization bubble in the particle-hole channel is negative, as opposed to the bubble in the particle-particle channel.

 The corresponding term in the Hamiltonian is
\begin{align}
\mathcal{H}_{c}=&U_c ~ (c^\dagger_{2, \alpha} c_{2,\nu})(c^\dagger_{1,\mu} c_{1,\beta})  \left[\frac92\delta_{\alpha\beta}\delta_{\mu\nu}+\frac12\vec\sigma_{\alpha\beta}\cdot\vec\sigma_{\mu\nu}\right],\nonumber\\
\label{ch_5}
\end{align}
where subindices 1 and 2 indicate that the corresponding momenta are near hot spots 1 and 2, and  the spin factors originate from
\begin{align}
&\(\vec \sigma_{\gamma\beta}\cdot \vec \sigma_{\alpha\delta}\)\(\vec \sigma_{\mu\gamma}\cdot \vec\sigma_{\delta\nu}\)\nonumber\\
&=\(\frac32\delta_{\alpha\beta}\delta_{\gamma\delta}-\frac12\vec\sigma_{\alpha\beta}\cdot \vec\sigma_{\gamma\delta}\)\(\frac32\delta_{\delta\gamma}\delta_{\mu\nu}-\frac12\vec\sigma_{\delta\gamma}\cdot \vec\sigma_{\mu\nu}\)\nonumber\\
&=\frac92\delta_{\alpha\beta}\delta_{\mu\nu}+\frac12\vec\sigma_{\alpha\beta}\cdot\vec\sigma_{\mu\nu}.
\label{wa_5}
\end{align}
Only the first, $\frac92\delta_{\alpha\beta}\delta_{\mu\nu}$ term is relevant to CDW instability as it renormalizes $\Delta^Q_k$, which acts between fermions near hot spots 1 and 2, with the same spin components (i.e., it is convoluted with $\delta_{\alpha \beta}$), therefore we can drop the $\frac12\vec\sigma_{\alpha\beta}\cdot\vec\sigma_{\mu\nu}$ component in Eq.\ (\ref{ch_5}).

The composite interaction $U_c$ can be approximated as a constant if
 the deviation of the fermionic momenta from corresponding hot spots (e.g., regions 1 and 2 in Fig.\ \ref{fig:0}) are smaller
  than the inverse spin correlation length $\xi^{-1}_s$. At larger deviations from hot spots, $U_c$ becomes function of momenta and decreases.

 We now turn to the calculation of the fermionic self-energy and pairing instability. At a first glance, an interaction with a positive (repulsive) $U_c$ cannot give rise to the pairing instability with sign-preserving gap between the regions 1 and 2. On a more careful look, however, we notice that
 %
the interaction in Eq.\ (\ref{ch_5})  is the one at  small momentum transfer  (both incoming and outgoing fermions are near the same hot spot), while to analyze CDW-mediated pairing (Fig.\ \ref{fig:sc}) and fermionic self-energy  (Fig.\ \ref{fig:5}(c)) one needs density-density interaction at momentum transfer approximately equal to the momentum difference between hot spots 1 and 2, namely, $Q_y=(0,Q)$. To obtain this
 interaction, we need to interchange one creation and one annihilation fermionic operator. Then we obtain
 from Eq.\ (\ref{ch_5})
\begin{align}
\mathcal{H}_{c}=-\frac{9U_c}{2}(c^\dagger_{2, \alpha} \delta_{\alpha\beta} c_{1,\beta})( c_{1,\mu}^\dagger\delta_{\mu\nu}c_{2,\nu}).
\label{ch_55}
\end{align}
 Viewed this way, the effective interaction is {\it attractive}
   and is capable to give rise to pairing.

\begin{figure*}
\includegraphics[width=1.8\columnwidth]{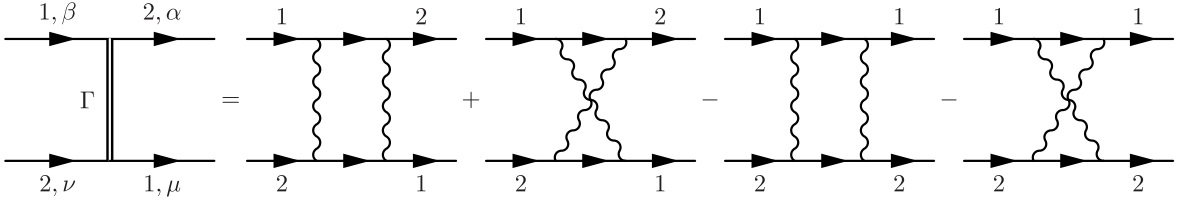}
\caption{The vertex function with momentum transfer near $Q_y$ in two-loop order (two spin-fluctuation propagators). Other two-loop diagrams (not shown) contain spin-fluctuation propagators with small momentum transfer and are irrelevant for our purposes. The charge component of this vertex function is $9U_c/2$. The opposite of this charge component (i.e., $-9U_c/2$) is the bare interaction in the charge channel at momentum transfer near $Q_x$ or $Q_y$. }
\label{fig:xx}
\end{figure*}
Another, more standard way  to verify that the density-density interaction with momentum transfer ${\bf Q}$ is attractive is to
  again extract it from the vertex function $\Gamma_{\alpha\mu,\beta\nu} ({\bf q},\Omega_m) $.  To second order in the spin-fluctuation propagator, there are two direct and two anti-symmetrized
   diagrams for $\Gamma_{\alpha\mu,\beta\nu} ({\bf q},\Omega_m)$ which contain spin propagators with momenta near ${\bf Q}_\pi$. We show them in
   Fig.\ \ref{fig:xx}.
    In evaluating these four diagrams, we additionally require that
      both spin-fluctuation propagators carry the same momenta, as only then one can cancel  extra power of ${\bar g}_s$.
        It is easy to show that only the fourth diagram (the one from the anti-symmetrized part) satisfies these conditions.
         This diagram is exactly the same as the one in Fig.\ \ref{fig:x}, but there is an extra minus sign in front of it.
           The evaluation of the diagram itself gives $-U_c$ because particle-hole bubble is
           negative. The extra $(-1)$ in front of this diagram cancels the overall minus sign. As a result, the charge component of the vertex function
            becomes
            %
\begin{align}
\Gamma^{c}_{\alpha\mu,\beta\nu}=
\frac{9 U_c}{2} \delta_{\alpha\beta}\delta_{\mu\nu}
\end{align}
where the spin structure is obtained the same way as in Eq.\ (\ref{wa_5}).
  Associating the charge component of the vertex function with the effective density-density interaction, we reproduce
   Eq.\ (\ref{ch_55}).

\begin{figure}[h]
\includegraphics[width=\columnwidth]{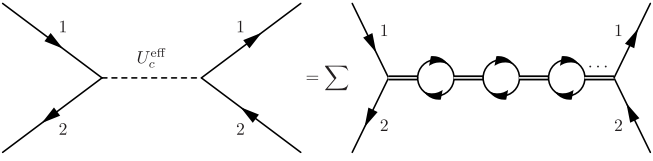}
\caption{The RPA diagrams for the dressed effective charge interaction $U^{\rm eff}_c$. Each double slid line is the "bare" $U_c$ -- the charge component of the vertex function at two-loop order. The dressed interaction $U^{\rm eff}_c$ can be viewed as charge fluctuation exchange (see text).}
\label{fig:xxx}
\end{figure}

The effective interaction $-(9U_c/2) c_{2,\alpha}^\dagger c_{1,\alpha} c_{1,\mu}^\dagger c_{2,\mu}$ is the {\it bare} interaction at momentum transfer $Q_y$, and in this respect $9U_c/2$ plays the same role as Hubbard $U$ played for our earlier derivation of spin-mediated interaction within RPA.
 Just like we did for spin case, we now dress interaction by summing up series of RPA diagrams
   (see
   Fig.\ \ref{fig:xxx}).
   This way we obtain fully renormalized (within RPA) effective interaction in the charge channel
 \beq
 U^{\rm eff}_c=\frac{9 U_c}{2} \frac{1}{1 - 9 U_c |\Pi_{c} ({\bf q},\Omega_m)|}
\label{ch_6}
\eeq
Expanding the polarization operator $\Pi_{c}$ near, say ${\bf Q} = Q_y$, we obtain
\beq
|\Pi_c ({\bf q}, \Omega_m)| = |\Pi (Q_y, 0)| - C_y ({\bf q} - {\bf Q}_y)^2 - \frac{|\Omega_m|}{\pi v^2_F}\frac{\alpha^2+1}{2\alpha}
\label{ch_3_a}
\eeq
Substituting this form into (\ref{ch_6}) we obtain the effective charge-mediate interaction  the same form as in Eq.\ (\ref{ch_8_1}) with
 \beq
 {\bar g}_c = \frac{1}{2 C_y},~~ \xi_c^{-2} = \frac{1-9U_c \Pi (Q_y,0)}{9U_c C_y},~~\gamma_c =  {\bar g}_c  \frac{1}{\pi v^2_F}\frac{\alpha^2+1}{\alpha}
  \label{ch_4_a}
  \eeq

 To proceed further
   we approximate the dynamical spin susceptibility $\chi_s ({\bf q}, \Omega_m)$ by its value at ${\bf q} = {\bf Q}_\pi$ and $\Omega_m=0$ and integrate over fermions within the momentum range of the width $\Lambda$ around hot spots.  Like we said above,  the approximation of $U_c$ by a constant is
     valid when momentum deviations from  a hot spot
       are at most of order $\xi^{-1}_s$, so  $\Lambda \xi_s$ is generally  of order one.
        We also assume for simplicity that CDW order parameter has a pure $d$-wave form, i.e., set
          our parameter $\alpha$ to be one.
          Within this last approximation,  the polarization operator has the same form between points  1-2 and 3-4.
   Evaluating the polarization operator $\Pi_c$
   we then find near ${\bf q} = {\bf Q}_y$~\cite{debanjan}
\begin{align}
|\Pi_c({\bf q})|=&\frac{\Lambda}{\sqrt{2}\pi^2v_F (1+ \lambda)}\left[1- {\tilde C}_y \frac{({\bf q}-{\bf Q}_x)^2}{\Lambda^2}\right]
\label{pi0p}
\end{align}
where  ${\tilde C}_y$ is of order one and we remind that
$\lambda = 3{\bar g}_s/(4\pi v_F \xi^{-1}_s)$ is the mass renormalization due to spin-fluctuation exchange.
 The variable $C_y$, which we introduced in (\ref{ch_3_a}) is related to ${\tilde C}_y$ as
\beq
C_y = {\tilde C}_y \frac{1}{\sqrt{2} \pi^2 v_F (1+\lambda)\Lambda}.
\eeq
Within the same approximation the composite interaction $U_c$ is given by
\beq
U_c = \left({\bar g}_s \xi^2\right)^2 |\Pi_c ({\bf Q_y})| = {\bar g}_s^2 \xi^3 \frac{\Lambda \xi}{\sqrt{2}\pi^2 v_F (1+ \lambda)}
\eeq
such that
\beq
9 U_c |\Pi_c ({\bf Q_y})| = (3 {\bar g}_s \xi)^2 \frac{(\Lambda \xi)^2 }{2\pi^4 v^2_F (1+ \lambda)^2}
\eeq
Using the condition $9 U_c |\Pi_c ({\bf Q_y})| \approx 1$, we eliminate unknown scale $\Lambda$ and obtain
\beq
C_y =  {\tilde C}_y  \frac{3 {\bar g}_s \xi^2_s}{2 \pi^4 v^2_F (1+\lambda)^2}
\eeq
Hence
\beq
{\bar g}_c = \frac{1}{2C_y} = {\bar g}_s \frac{3\pi^2}{16 {\tilde C}_y} \left(\frac{1+\lambda}{\lambda}\right)^2 \approx 2 {\bar g}_s
\frac{1}{{\tilde C}_y} \left(\frac{1+\lambda}{\lambda}\right)^2
\eeq
We see that within this approximation the ratio ${\bar g}_c/{\bar g}_s$ depends on the value of dimensionless parameter ${\tilde C}_y$.
To obtain this parameter one needs to know more precisely system behavior at energies comparable to $\Lambda$.  Still, if ${\tilde C}_y (\lambda/(1+\lambda))^{2} \leq 1$, then ${\bar g}_c \geq 2 {\bar g}_s$, in which case superconducting $T^{\rm ch}_c$ from the exchange of near-critical charge fluctuations
well may exceed $T^{\rm sp}_c$ from the spin-fluctuation exchange.  This is the central result of this section.

A more quantitative analysis requires extensive numerical calculations and is beyond the scope of this work. We also emphasize that there is no known controllable procedure of the derivation of the effective interaction mediated by  near-critical collective bosonic fluctuations, the RPA which we used
 is an uncontrollable approximation. And we also recall that the composite interaction $U_c$ (the convolution of two fermionic propagators and two spin propagators) does depend on external momenta and frequency,
  and already the calculation of $T_{\rm cdw}$ requires one to solve integral equation for momentum and frequency dependent full $\Delta^Q_k$.

\section{ Shrinking of a superconducting dome in a magnetic field}

 Finally, we discuss the issue of how $T^{\rm ch}_c$, mediated by collective degrees of freedom, evolves in the presence of an external magnetic field.  For definiteness, we focus on the role of near-critical charge fluctuations and neglect the contribution to $T_c$ from spin fluctuations.

 It has been found experimentally~\cite{ramshaw, ramshaw_science} that in the presence of a magnetic field  $H$  superconducting $T_c$, viewed as a function of doping, splits into two domes, and the one at higher doping is centered at or very near CDW QCP. As $H$ increases,
 the maximum of $T_c$ in this dome is somewhat reduced, but, most notably,  the width of the dome shrinks, i.e.,  superconductivity get progressively confined to a CDW QCP.

\begin{figure}
\includegraphics[width=\columnwidth]{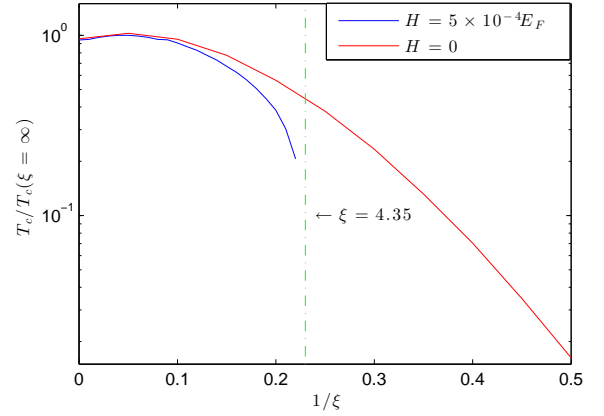}
\caption{The behavior of $T^{\rm ch}_c$'s as a function of $1/\xi_c$ with and without an external field $H$, obtained by explicitly
 solving Eq.\ (\ref{sc333}). Without a magnetic field  $T^{\rm ch}_c$  decreases  as $\xi_c$ becomes finite and at small enough $\xi_c$ crosses over from quantum-critical to BCS-like behavior [see Eq. (\ref{ch_n3})]. At a finite field, $T^{\rm ch}_c$ at $\xi_c=\infty$ is somewhat reduced, but, most important,
  $T^{\rm ch}_c$ now vanishes at a
 finite $\xi_c^{\rm cr}$. In numerical calculations we used
 $\bar g_c=0.75 E_F$, $\alpha=0.076$, and $\mu_BH=5\times 10^{-4} E_F$ (for $E_F = 1~{\rm eV}$, this $H$ is $\sim$10 Tesla). For these parameters,
 $\xi_c^{\rm cr} =4.35/k_F$. The critical $\xi_{c}^{\rm cr}$ is well described by  Eq.\ (\ref{yw74}) --
  plugging $\xi_c^{\rm cr}$ into this equation gives 1.2 in the right hand side, close enough to the actual 1. }
\label{H}
\end{figure}

 We show that this behavior is reproduced within a quantum-critical CDW pairing scenario. The argument is rather straightforward -- right at CDW QCP,
  $T^{\rm ch}_c$ is set by charge-fermion coupling ${\bar g}_c$, and to reduce $T^{\rm ch}_c$ one would need to apply a rather strong magnetic field $\mu_B H \sim {\bar g}_c$.  Away from CDW QCP, in the FL regime, $T^{\rm ch}_c$ is reduced and eventually follows BCS formula. In the latter case, a much weaker $\mu_B H$ is needed to kill superconductivity.

 To see how this works in practice, we solved for charge-mediated $T^{\rm ch}_c$ at a finite charge correlation length $\xi_c$ by assuming that the dominant effect of the field is Zeeman splitting of fermionic energies in the particle-particle bubble.  Within this approximation, the linearized integral equation for the pairing vertex function $\Phi (\omega_m,k_{\parallel},)$ (the one which has a solution at $T=T^{\rm ch}_c$) is
  \begin{align}
&\Phi(\omega_m,k_{\parallel})\nonumber\\
=&\frac{\bar g_c T}{v_F}\sum_{m'}\int \frac{dk'_{\|}}{2\pi}\frac{\Phi(\omega'_m,k'_{\parallel})\sgn(\omega_m')}{\omega_m'-i\Sigma(\omega_m', k'_{\|})-i\mu_B H}\nonumber\\
&\times\frac{1}{k_\|^2+k_\|^{\prime2}-2
\beta
 k_\|k_\|'+\gamma_c|\omega_m-\omega_m'|+\xi_c^{-2}}\nonumber\\
=&\frac{\bar g_c T}{v_F}\sum_{m'}\int \frac{dk'_{\|}}{2\pi}\frac{\Phi(\omega'_m,k'_{\parallel})~|\omega_m'-i\Sigma(\omega_m', k'_{\|})|}{[\omega_m'-i\Sigma(\omega_m', k'_{\|})]^2+(\mu_B H)^2}\nonumber\\
&\times\frac{1}{k_\|^2+k_\|^{\prime2}-2\beta k_\|k_\|'+\gamma_c|\omega_m-\omega_m'|+\xi_c^{-2}},
\label{sc333}
\end{align}
where the self-energy is given by Eq. (\ref{sigmaxy_1}).

In the FL regime, when $\lambda_c = {\bar g}_c/(2\pi v_F \xi^{-1}_c) \leq 1$, $\Phi (k_{\parallel},\omega_m)$ can be, to logarithmical accuracy,  approximated by a constant $\Phi$, and Eq. (\ref{sc333}) reduces to

\begin{align}
\Phi=\frac{\lambda_c}{1+\lambda_c}\log\frac{\omega_{\rm cf}}{(T^2_c+H^2)^{1/2}}\Phi.
\end{align}
where $\omega_{\rm cf} \sim {\bar g}_c/\lambda^2$ has been introduced in (\ref{ch_n2}).
The superconducting $T^{\rm ch}_c$ becomes zero at a critical $\lambda_{c}^{\rm cr}$, given by
\begin{align}
\frac{\lambda_c^{\rm cr}}{1+\lambda_c^{\rm cr}}\log\frac{\omega^{\rm cr}_{\rm cf}}{H}=1,
\label{yw74}
\end{align}
or, with logarithmical accuracy, at $\lambda_c^{\rm cr} \sim 1/\log({{\bar g}_c/\mu_B H})$.  At smaller $\lambda_c$, i.e., at larger deviations from CDW QCP, there is no charge-mediated superconductivity.

\begin{figure}[h]
\includegraphics[width=\columnwidth]{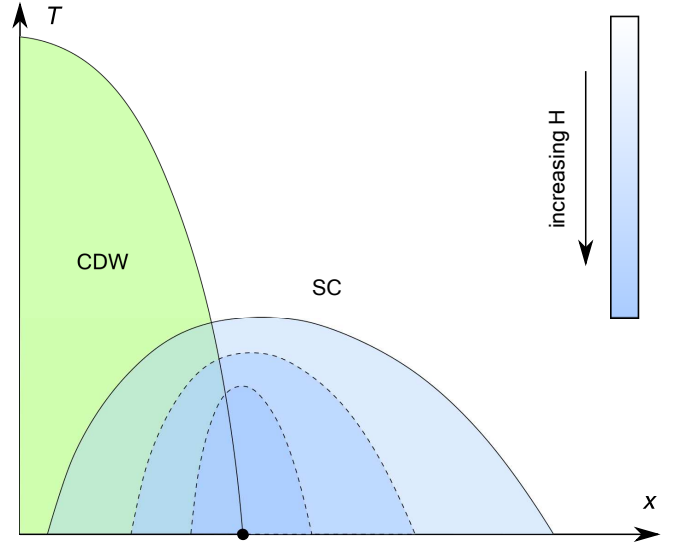}
\caption{The variation of the onset temperature of superconducting order mediated by  near-critical charge fluctuations in the presence of an  external field $H$. As our numerical results show (Fig.\ \ref{H}), the range of the superconducting dome shrinks as magnetic field increases.}
\label{phases_H}
\end{figure}

 We solved the gap equation numerically and obtained $T^{\rm ch}_c$ as a function of $\lambda_c$ and $\mu_B H$. We plot the results in Fig.\ \ref{H},
  and
  present the corresponding phase diagram schematically in Fig.\ \ref{phases_H}.  We see that, indeed, superconducting dome gets sharper in the field, i.e., charge-mediated superconductivity gets progressively confined to CDW QCP.  We did not do calculations on the other (ordered) side of CDW QCP, but by generic reason we expect a similar shrinking of $T_c$ range.
The shrinking of $T_c$ range with increasing field is fully consistent with the experimental data~\cite{ramshaw, ramshaw_science}.

\section{Conclusion}

Motivated by the observation of a static charge order in the cuprates and the enhancement of $T_c$ at its onset, we studied in this work
the pairing mediated by charge fluctuations around the quantum-critical point towards an incommensurate charge order with  momentum   $Q_x = (Q,0)$ or $Q_y = (0,Q)$.  Our main goal was to understand whether charge-mediated pairing near a CDW QCP yields $T_c$ comparable to that obtained
from spin-fluctuation exchange.

We first considered a semi-phenomenological charge-fermion model in which hot fermions (the ones at the FS, connected by $Q_x$ or $Q_y$) interact by exchanging soft collective excitations in the charge channel.
We obtained  bosonic and fermionic self-energies in the normal state and used them as inputs
 for the analysis of the quantum-critical pairing problem. We found, in agreement with earlier works~\cite{italians}  that the charge-mediated pairing interaction is attractive in both $d$-wave  and $s$-wave channels. The $d$-wave pairing becomes more favorable once we
  include other contributions to the pairing interaction from, e.g.,  antiferromagnetic spin fluctuations.
 We found that the critical temperature $T_c$ scales with the charge-fermion coupling constant $\bar g_c$, and that fermions
 from NFL regime very near a hot spot and from a FL region further away from a hot spot
 contribute to the pairing. In this respect, pairing near a CDW QCP is similar to the pairing by spin fluctuations near a SDW QCP.  We obtained the value of $T^{\rm ch}_c/{\bar g}_c$ numerically.

  We next considered the microscopic model in which spin fluctuations emerge at higher energies than charge fluctuations and are  therefore the primary collective degrees of freedom.
     Charge fluctuations emerge at smaller energies as composite fields, made out of pairs of spin fluctations.  Within this model, we were able to express charge-fermion coupling ${\bar g}_c$ via the underlying spin-fermion coupling ${\bar g}_s$ and relate $T^{\rm ch}_c$ due to charge-fluctuations near a CDW QCP to $T^{\rm sp}_c$  due to spin fluctuations.  We found that, at least within RPA, $T^{\rm ch}_c$ due to charge fluctuations is comparable to
      $T^{\rm sp}_c$ that due to spin fluctations and may even exceed it, i.e., superconducting $T_c$ does get a substantial enhancement near a CDW QCP.

  Finally, we analyzed the behavior of charge-mediated $T_c$ in the presence of a magnetic field and found that the dome of $T^{\rm ch}_c$ around a CDW QCP indeed shrinks as magnetic field increases, because a field destroys superconductivity  faster in non-critical regime than in the quantum-critical regime and hence enhances charge-fluctuation component of $T_c$ near a CDW QCP.

   This result  and the one that the contribution to $T_c$ from critical charge fluctuations  can be larger than the contribution to $T_c$  from non-critical  spin-fluctuations, despite that charge fluctuations are by themselves made out of spin fluctuations, may explain the experimental observation that in a magnetic field $T_c$ gets progressively confined to the doping range around the doping at which charge order likely emerges at $T=0$.

   \begin{acknowledgments}
We thank D. Agterberg, E. Berg, R. Fernandes, S. Kivelson, S. Lederer, Y. Shattner, and B. Shklovskii for fruitful discussions. The work was supported by NSF/DMR-1523036.
 \end{acknowledgments}

\end{document}